# Syngas Production from Propane using Atmospheric Non-Thermal Plasma

F. Ouni, A. Khacef* and J. M. Cormier

GREMI - Polytech'Orléans, 14 rue d'Issoudun, BP 6744, 45067 Orléans Cedex 2, France

**Abstract**

Propane steam reforming using a sliding discharge reactor was investigated under atmospheric pressure and low temperature (420 K). Non-thermal plasma steam reforming proceeded efficiently and hydrogen was formed as a main product ($H_2$ concentration up to 50%). By-products ($C_2$-hydrocarbons, methane, carbon dioxide) were measured with concentrations lower than 6%. The mean electrical power injected in the discharge is less than 2 kW. The process efficiency is described in terms of propane conversion rate, steam reforming and cracking selectivity, as well as by-products production. Chemical processes modelling based on classical thermodynamic equilibrium reactor is also proposed. Calculated data fit quiet well experimental results and indicate that the improvement of $C_3H_8$ conversion and then $H_2$ production can be achieved by increasing the gas fraction through the discharge. By improving the reactor design, the non-thermal plasma has a potential for being an effective way for supplying hydrogen or synthesis gas

## 1. Introduction

The increasing concern environmental and pollution problems and the fossil resources evolution implied a large amount of effort in the scientific world to develop alternative and renewable technologies for energy and electricity production. Nowadays, fossil fuels were the first energy source but their burning processes induce the emission of a large amount of soot, carbon dioxide and nitrogen oxides responsible for the greenhouse effect. In this context, the hydrogen as a clean fuel appears to be the best future solution for fuel cells and automotive applications [1, 2].

Synthesis gas or syngas (mixture of hydrogen and carbon monoxide) are used as a major intermediary for the production of pure hydrogen or other chemical compounds. About 50% of hydrogen is produced from natural gas by several methods, e.g. partial oxidation with oxygen (PO), steam reforming (SR), steam reforming with oxygen (SRO), CO2 reforming (CDR) and $CO_2$ reforming with oxygen (CDRO) [3]. One of the most attractive technologies seems to be the SRO, which is also known as auto-thermal reforming. The main reasons for recommending this technology over the other variants are the low energy requirements, the high space velocity (at least one order of magnitude relative to traditional SR), a lower process temperature than PO and the $H_2$/CO ratio is easily regulated by the inlet gas ratio [4-6]. The catalysts commonly used are based on transition metals (e.g. Ni, Rh, Pt, Ir) supported on metal oxides, such as $Al_2O_3$, $CeO_2$, MgO, $TiO_2$ or rare-earth oxides.

Although presenting high performances, the chemical auto-thermal reformers present several shortcomings such as large size of the equipment, high investment and exploitation costs, limitations on rapid response, extreme operating conditions limiting the reactor lifetime, catalyst sensitivity to poisons (coking and deactivation) [7-10], safety, and operability.

The conventional reformers allowing syngas production are based on steam reforming of hydrocarbons [3] following the reaction (1):

*Corresponding author. Tel.: +33 2 38 49 48 75, Fax.: +33 2 38 41 71 54.
 E-mail address: ahmed.khacef@univ-orleans.fr

$$C_nH_m + nH_2O \rightarrow \left(n + \frac{m}{2}\right)H_2 + nCO \qquad (1)$$

The steam reforming of hydrocarbons is strongly endothermic ($\Delta H^0$=498 kJ.mol$^{-1}$ for $C_3H_8$) and the process requires high temperatures (700-1200 K) to be carried out. Therefore, finding a steam reforming process operating under "sweet" conditions is expected to be a challenge from the viewpoint of industrial application as well as fundamental science.

In plasma reformer process, electrical discharge could be used alone [11-16] or combined with catalyst [17-23]. At present, plasma reformer gives an attractive alternative for hydrogen and syngas production. Recently, non-thermal plasma (NTP) or cold plasma technology has been developed for chemical applications (metallurgy, microelectronics, environmental applications) with both high efficiency and selectivity for chemical reactions. Investigations of these fields have been made worldwide by corona discharge, dielectric barrier discharges (DBD) or gliding arc at atmospheric pressure and ambient temperature. In NTP, the electrons are energetic whereas ions and neutral gas are near ambient temperature. Typical electron temperature ($T_e$) in such plasmas is of order of few electron volts (eV), which is sufficient to breakdown molecules and to produce highly active species: radicals, excited atoms, ions and molecules as well as electrons. As example, in gliding discharge reactor, the on-axis temperature is in the range of 5000 K while the mean gas temperature remains around ambient temperature. Such non-equilibrium plasma could replace catalyst and enhance chemical processes through highly active species: radicals, excited atoms, ions and molecules as well as electrons. Plasma reactors could represent an incisive approach by their simplicity, high conversion efficiencies, fast response time, compactness, and low energetic cost.

The present work is a continuation of our previous studies regarding the syngas production by steam reforming of hydrocarbons [12-15]. In the set of experiments presented in this paper, propane steam reforming was investigated using a sliding discharge reactor (SDR) operating at atmospheric pressure and inlet gas temperature of about 150°C without the use of any catalyst. The process efficiency is described in terms of $C_3H_8$ conversion rate, $H_2$ production, steam reforming/cracking selectivity, and by-products production. Experimental results were compared with a thermodynamic model describing the chemical evolution of the system.

## 2. Experimental

The experimental arrangement that was used in this work is shown on figure 1. Before injection in the plasma reactor, steam and propane are mixed in a mixing chamber using calibrated high-precision mass flowmeters. Total gas flow rate ranging between 80 and 110 L/min are used. All experiments were conducted at atmospheric pressure. The inlet gas temperature was fixed at about 150°C by using two heated lines. After plasma treatment the exhausts gas was passed through condensers.

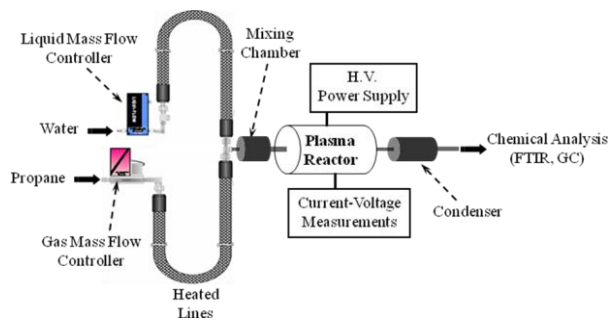

Figure 1: Experimental arrangement (FTIR: Fourier Transform Infrared Spectroscopy, GC: Gas Chromatography)

The NTP reactor used in that study is a SDR type (sliding discharge reactor) in a cylindrical configuration (figure 2). SDR was described in details previously [12] and is briefly described here for clarity. It consists on three copper anodes arranged around a single tungsten cathode inserted in quartz tube. Discharges are ignited between electrodes and then pushed by the gas flow. A magnet was inserted in the reactor in order to produce a rotating effect in the discharge region. The discharge column is a plasma string, with a visible diameter less than one millimetre that slides in the gas flow and the magnetic field region. As shown on figure 2, the plasma string performs a helix movement and looks like a wrapped wire around the cathode.

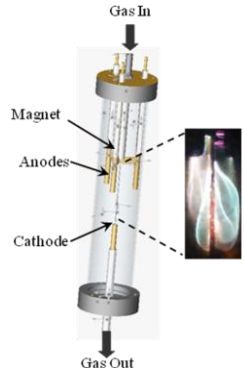

Figure 2: Schematic of the SDR and photography of the discharge

The SDR was driven by a three-channel power supply device. Details of this home made high-voltage generator are presented in [8]. Briefly, the electrical supply of the reactor is a three channels (15 kV, 50 Hz) one and is used to power up three cascading discharges. One channel consists of two transformers with rectifiers that allow two running phases: the ignition at high voltage and low current intensity and a complementary energy supply with higher current. Current and voltage measurements were achieved using a Hall Effect probe and voltage divider (ratio of 0.01), respectively. The output signals were transmitted to a transient digitizer (Tektronix TDS 3034B) interfaced with a personal computer. Typical voltage and current waveforms for one of the three discharges are shown on figure 3. For clarity, the high voltage is plotted as a negative signal. As it can be seen on this figure, the discharge behaviour is not definitely periodic due to the instability in the growing discharge.

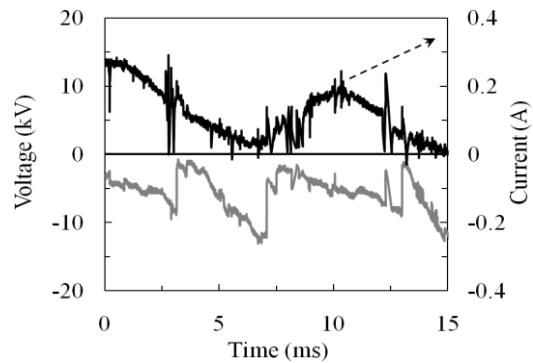

Figure 3: Typical current and voltage waveforms

The mean electrical power averaged over a large number of discharges is calculated from the current and the voltage waveforms by:

$$P = \frac{1}{T} \int_0^T u(t).i(t).dt \qquad (2)$$

In this type of reactors, plasma can sweep a large part of the inlet gas and maintains its non-equilibrium behaviour. The force acting on the lengthening discharge column is proportional to the product between the current and the magnetic-field strength. This force produces a rapid lengthening of the discharge column and then increases the plasma resistance. In that case, a self-limitation of the current intensity is produced. In the usual sliding discharges, the plasma thermalization was avoided by using external current limitation.

The outlet gas pass through condensers and then was analysed and quantified using two techniques: micro gas chromatography (µGC, Varian CP2003-P) and Fourier Transform Infra Red spectroscopy (FTIR, Nicolet Magna-IR 550 series II). The µGC analyser was equipped with MolSieve 5Å and PoraPlot Q columns. Both columns were equipped with thermal conductivity detector (TCD) calibrated with standards of known composition. The main gas components identified were $H_2$, $CO$ and no transformed $C_3H_8$. Small amounts of $CO_2$, $C_2$-hydrocarbons ($C_2H_2$ and $C_2H_4$) and $CH_4$ have been detected with concentrations lower than 6% in all studied cases. The $H_2O$ concentration at reactor output was estimated from the C, H and O balances.

## 3. Results and discussions

Propane steam reforming (SR) process described by the reaction (3) is strongly endothermic ($\Delta H^0 = 498$ kJ.mol$^{-1}$).

$$C_3H_8 + 3H_2O \rightarrow 7H_2 + 3CO \tag{3}$$

Results of thermodynamic calculations shown on figure 4 demonstrate that the SR reaction requires high temperature (700-1200 K) to be achieved. $C_3H_8$ SR takes place at temperature higher than 600 K. Total propane conversion is reached at about 800 K. Thus, higher temperature is necessary to activate $C_3H_8$ molecule.

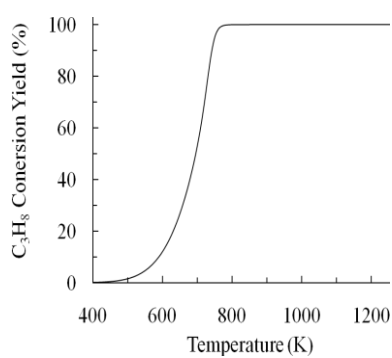

Figure 4: Conversion rate of the SR of $C_3H_8$ as a function of the temperature

(Thermodynamic calculations)

In conventional catalytic technology and even in plasma reformer, SR reaction suffered from competitiveness with the cracking reaction (CR) described by the general reaction (4):

$$C_nH_m \rightarrow \frac{m}{2}H_2 + nC \tag{4}$$

In the case of $C_3H_8$ cracking reaction using conventional catalytic scheme, coke deposition is observed [8] even when the experiments were carried out in the region expected from the equilibrium to be carbon free. In plasma processing, carbon deposit on the walls of the reactor and on electrodes could be a serious problem with a consequence of decreasing the system efficiency [11-21]. However, one can note that high hydrogen purity can be obtained through that reaction avoiding CO and $CO_2$ production.

In the following, experimental data obtained after condensation of water (dry conditions) are presented in terms of $C_3H_8$ conversion rate, SR - CR selectivity, and species production as a function of the plasma reactor inlet parameters (steam to propane ratio and total flow rate). Figure 5 shows an example of data in dry conditions as a function of inlet steam to $C_3H_8$ concentrations ratio for 80 L/min total flow rate. Analysis show that the concentration of the main species depends strongly on the steam to propane inlet ratio. Increasing the $H_2O$ to $C_3H_8$ ratio leads to lower the amount of the no-transformed $C_3H_8$. The maximum $H_2$ concentration (50%) was obtained for the higher $H_2O$ to $C_3H_8$ ratio (about 19). Compared to previous studies on $CH_4$ steam reforming [12], the plasma $C_3H_8$ SR produced a large amount of $C_2$-hydrocarbons and $CH_4$ (concentrations up to 8% and 2.5%, respectively) for the lower $H_2O$ to $C_3H_8$ ratio (about 4) as shown on figure 6. It seems that the $C_3H_8$ conversion is initiated by its decomposition into lower hydrocarbons, then the SR reaction occur with these lower hydrocarbons [24]. $C_3H_8$ SR generates a large amount of $CO_2$ molecules (up to 4% for $H_2O$ to $C_3H_8$ ratio of about 4). However, as shown on figure 6, using a sliding discharge reactor $CO_2$ production can be prevented for molar steam to propane ratio higher than 10. This behaviour is quite similar whatever the total gas flow rate ranged between 80 and 110 L/min.

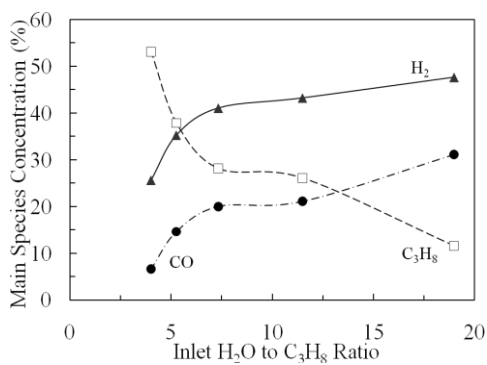

Figure 5: Main species concentration as a function of the inlet steam to propane ratio for total flow rate of 80 L/min.

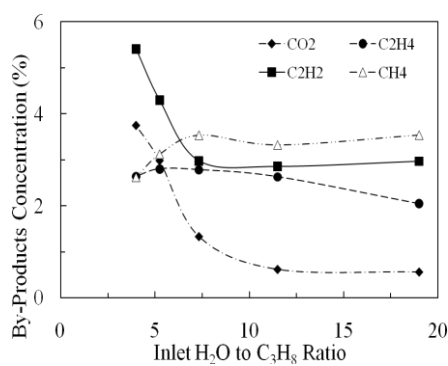

Figure 6: By-products concentration as a function of the inlet steam to propane ratio for total flow rate of 80 L/min.

Taking into account the main products identified and quantified at the outlet of the SDR by chromatography and FTIR diagnostics, the characteristic parameters of the process such as the $C_3H_8$ conversion rate and the SR-CR selectivity were calculated using the steam reforming reaction (1), the main possible cracking reactions (5), (6), and (7), and the water gas shift reaction (8):

$$C_3H_8 \rightarrow 3H_2 + C_2H_2 + C \qquad \Delta H^0 = 313 \text{ kJ.mol}^{-1} \qquad (5)$$

$$C_3H_8 \rightarrow 2H_2 + C_2H_4 + C \qquad \Delta H^0 = 156 \text{ kJ.mol}^{-1} \qquad (6)$$

$$C_3H_8 \rightarrow 2H_2 + CH_4 + 2C \qquad \Delta H^0 = 30 \text{ kJ.mol}^{-1} \qquad (7)$$

$$CO + H_2O \rightarrow H_2 + CO_2 \qquad \Delta H^0 = -41 \text{ kJ.mol}^{-1} \qquad (8)$$

Species amounts consumed by particularly reaction were defined by the parameters $\alpha$, $\beta$, $\chi$, $\delta$, and $\varepsilon$ as follow:

- $\alpha$ is the $C_3H_8$ amount consumed via the SR reaction (1),
- $\beta$, $\chi$, and $\delta$ are the $C_3H_8$ amounts consumed via the CR (5), (6) and (7), respectively,
- $\varepsilon$ is the CO amount consumed via the WGS reaction (8).

For an initial $C_3H_8$ mole number equal to 1, the total amount of propane converted will be equal to $(\alpha + \beta + \chi + \delta)$. The $\alpha$, $\beta$, $\chi$, $\delta$, and $\varepsilon$ parameters were calculated as a function of species concentrations given by experimental analysis obtained by gas chromatography. Results are given by the following formula where the brackets design the specie concentrations.

$$\alpha = \chi \frac{([CO_2] + [CO])}{3[C_2H_4]} \qquad (9)$$

$$\beta = \chi \frac{[C_2H_2]}{[C_2H_4]} \qquad (10)$$

$$\delta = \chi \frac{[CH_4]}{[C_2H_4]} \qquad (11)$$

$$\varepsilon = \chi \frac{[CO_2]}{[C_2H_4]} \qquad (12)$$

$$\chi = \frac{[H_2]}{(9[H_2] - 7)\frac{[CO_2] + [CO]}{3[C_2H_4]} + (2[H_2] - 2)\left(\frac{[CH_4] + [CO_2]}{[C_2H_4]} + 1\right) + (3[H_2] + 3)\frac{[C_2H_2]}{[C_2H_4]}} \qquad (13)$$

The evolution of $C_3H_8$ conversion rate as a function of the feeding parameters of the reactor (flow rate and inlet steam/propane ratio) is shown on figure 7. The increasing of the initial propane amount decreases the propane conversion rate. At an initial propane amount of about 5%, conversion rate was about 26%. This value decreases to 8% using an inlet propane amount equal to 20%. The obtained results exhibit low values of the propane conversion rate (between 7 and 30%). These results were explained previously [12, 13] and are attributed to the design of the plasma reactor itself. In fact, we demonstrated that only a part of the injected gas mixture passed through the active region of the plasma reactor.

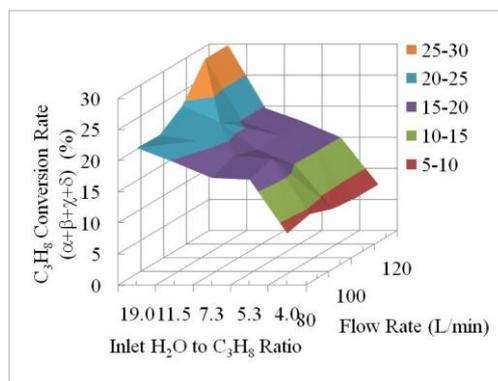

Figure 7: Calculated C₃H₈ conversion rate (α+β+χ+δ) as a function of the inlet steam to propane ratio.

The steam reforming selectivity, defined by the expression (14), was calculated for different experimental conditions.

$$C_3H_8 \text{ SR Selectivity} = \frac{\alpha}{\alpha + \beta + \chi + \delta} \tag{14}$$

Results of these calculations as a function of flow rates and inlet steam/propane ratio are shown on figure 8. For the total flow rate in the range 80-120 L/min, the maximum value of the $C_3H_8$ SR selectivity is between 50 and 60% showing that the cracking reaction was promoted even for high $H_2O$ concentration.

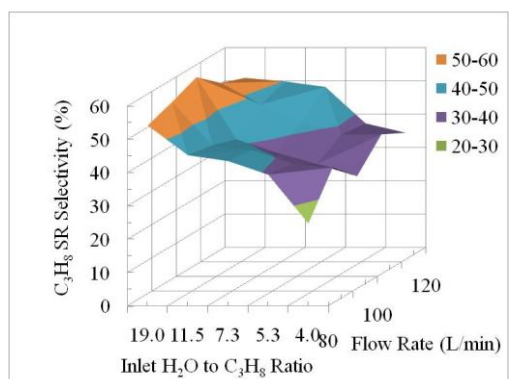

Figure 8: Calculated C₃H₈ steam reforming selectivity as a function of the inlet steam to propane ratio.

The plasma chemical processes, occurring in the SDR, was simulated by means of Chemical Workbench code [25, 26] (kinetic technologies, Russia) using a thermodynamic equilibrium reactor (TER) model. The TER is designed for the calculation of the chemical equilibrium of multi-components heterogeneous system. This thermodynamic system is considered as self-contained and closed. In this system, a state of thermodynamic equilibrium is achieved by internal chemical and phase transformations. It means that the system is under mechanical and energetic equilibrium. It is proposed also that the system under investigation is heterogeneous and consists of several uniform phases. The TER code uses the common principle of maximum entropy for the calculation of chemical and phase composition. In accordance with this principle, the equilibrium state is characterized by a uniform distribution of thermodynamic parameters in system volume and the chemical composition corresponds to maximum of probability of energetic levels distribution for macro particles. The method takes into account 268 atomic and ionic species that could appear in the C-H-O system. Calculations are made at atmospheric pressure for various temperatures.

We consider that the reactions occur in the discharge region and only a part of inlet gas mixture is effectively processed by the plasma. The model is presented schematically on figure 9. The inlet gas stream with a flow rate Q is divided in two parts and the γ-factor represents the amount of gas passed through the discharge region. We assume that at the outlet of the reactor, the both streams are perfectly mixed and the resulting stream is in thermodynamic equilibrium. γ-factor and plasma temperature T are the fitting parameters of the model.

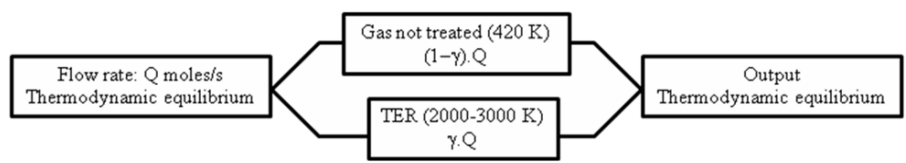

Figure 9: Simplified scheme of the model.

Calculated chemical equilibrium composition was plotted as a function of temperature for different $C_3H_8$-$H_2O$ systems at flow rate at about 80 L/min. Results are shown on figure 10. These data show a relative stabilization of main species concentration in the temperature range 2000-3000 K, that correspond to the temperature of the plasma column[27,28].

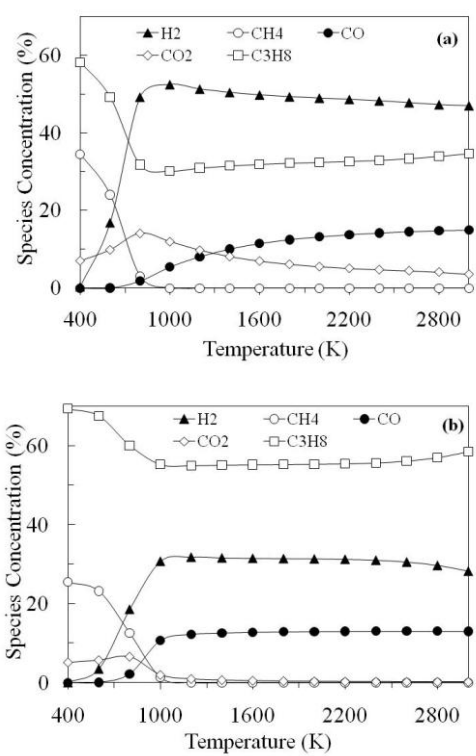

Figure 10: Computed species concentration at thermodynamic equilibrium as a function of temperature for different $H_2O$ - $C_3H_8$ systems at total flow rate and γ-factor of about 80 L/min and 50%, respectively: (a) $H_2O/C_3H_8$ = 19 and (b) $H_2O/C_3H_8$ =4.

The comparison of the experimental and theoretical data was made by adjusting the fitting parameters γ-factor and plasma column temperature T for each gas mixture studied (different inlet steam to carbon ratios). The best fits of experimental data was obtained with a γ-factor ranged between 40 and 55% and a plasma column temperature of about 2000 K. Figure 11 plots predicted and measured concentrations of major products and by-products in terms of inlet $H_2O$ to $C_3H_8$ ratio for a total flow rate of 80 L/min. Agreement between predicted and experimental results is satisfactory (agreement is within 10-15%).

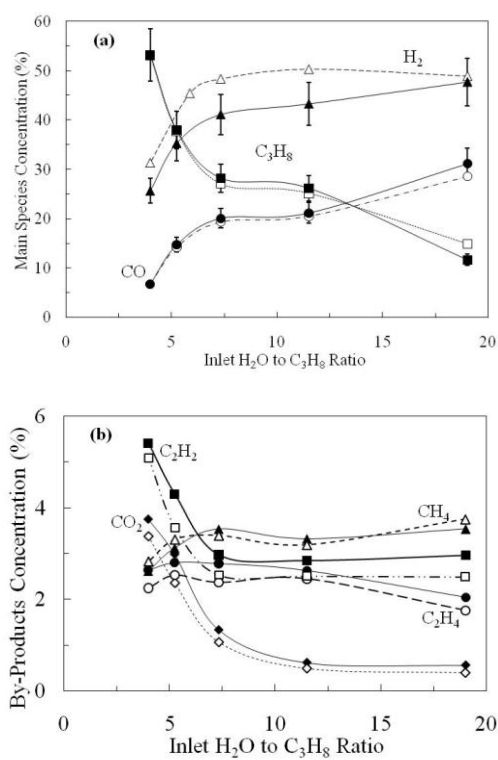

Figure 11: Experimental (full line) and calculated (dashed line) concentrations at thermodynamic equilibrium as a function of the inlet steam to propane ratio at 2000 K and total flow rate of 80 L/min: (a) main species and (b) by-products.

Plasma processing cannot be fully explained by thermodynamic equilibrium. However, one must note that the corresponding model demonstrates the possibility to obtain higher conversion rates by increasing the γ-factor. However new investigations are needed to find more adequate model for non-thermal plasma chemistry.

## 5. Conclusion

The conversion of propane into hydrogen and syngas was experimentally investigated in a sliding discharge reactor under conditions of low temperature (420 K), atmospheric pressure, and input power lower than 2 kW without the use of a catalyst. Interesting preliminary results were obtained.

Propane conversion, steam reforming selectivity, cracking selectivity, and hydrogen production depend strongly on the inlet steam to carbon ratio and the total gas flow rate. The obtained results exhibit low values of the propane conversion rate (7 - 30%). In any studied cases, the main products of the plasma treatment are hydrogen (50%) and carbon monoxide (up to 20%). By-products such as $C_2$-hydrocarbons ($C_2H_2$, $C_2H_4$), methane and carbon dioxide are presents at the outlet of the discharge reactor at relatively high concentrations (8% and 2.5%, respectively) while the amounts of these species were less than 1% when methane was used. Carbon deposition on the electrodes and reactor walls was observed even at high molar steam to carbon ratio causing destabilization of the discharge.

The plasma chemical processes occurring in the sliding discharge reactor SDR were simulated by means of Chemical Workbench code using a thermodynamic equilibrium reactor (TER) model. The TER is designed for the calculation of the chemical equilibrium of multi-components heterogeneous system. Thermodynamic calculations were performed by varying the plasma temperature and the amount of gas passed through the discharge region (γ-factor).

The best fit of experimental data was obtained with a γ-factor ranged between 40 and 55% and a plasma temperature of about 2000 K. These results show that only a part of inlet compounds was processed by the plasma reactor and could explain the low propane conversion rates (30%) obtained in that study. Further investigations must be performed in order to improve the plasma string volume by improving the plasma reactor configuration.